\documentclass[12pt,journal]{IEEEtran}

\ifCLASSINFOpdf
   \usepackage[pdftex]{graphicx}
   \usepackage{epstopdf}
\else
   \usepackage[dvips]{graphicx}
\fi
\usepackage{times}
\usepackage{dsfont}
\usepackage{cite}
\usepackage{indentfirst}
\usepackage{multirow}
\usepackage{amsmath}
\usepackage{amsfonts}
\usepackage{amssymb}
\usepackage{subfigure}
\usepackage{marvosym}
\usepackage{hyperref}
\usepackage{tikz}
\usepackage{booktabs}
\usepackage{float}
\usepackage{enumerate}
\usepackage{eqparbox}
\usetikzlibrary{shapes,arrows}

\begin{document}

\title{The Problem of Peak-to-Average Power Ratio in OFDM Systems}
\author{
\IEEEauthorblockN{ Martha C. Paredes Paredes{\small$^{\#1}$} \quad and \quad  M.~Julia Fen\'{a}ndez-Getino Garc\'{i}a{\small$^{\#2}$}} \\
$^{\#}$\textit{Department of Signal Theory and Communications \\
Universidad Carlos III de Madrid}\\
Madrid, Spain\\
\fontsize{11}{11}\selectfont\ttfamily\upshape
e-mail: $^{1}$mcparedes@tsc.uc3m.es  \quad $^{2}$mjulia@tsc.uc3m.es
}
\maketitle
\begin{abstract}
Orthogonal Frequency Division Multiplexing (OFDM) is widely used in many digital communication systems due to its advantages such us high bit rate, strong immunity to multipath and high spectral efficiency but it suffers a high Peak-to-Average Power Ratio (PAPR) at the transmitted signal. It is very important to deal with PAPR reduction in OFDM systems to avoid  signal degradation. Currently, the PAPR problem is an active area of research and in this paper we present several techniques and that mathematically analyzed. Moreover their advantages and disadvantages have been enumerated in order to provide the readers the actual situation of the PAPR problem. 
\end{abstract}

\IEEEpeerreviewmaketitle
\section{Introduction}

\IEEEPARstart{O}{rthogonal} Division Multiplexing (OFDM) is a technique widely used in many digital communication systems such us Digital Television (DTV), Digital Audio Broadcasting (DAB), Terrestrial Digital Video Broadcasting (DVB-T), Digital Suscriber Line (DSL) broadband internet access, standards for Wireless Local Area Networks (WLANs), standards for Wireless Metropolitan Area Networks (WMANs), and 4G mobile communications. It has many advantages such us high bit rate, strong immunity to multipath and high spectral efficiency. However, one of the most serious problems is the high Peak-to-Average Power Ratio (PAPR) of the transmitted OFDM signal, since this large peaks introduce a serious degradation in performance when the signal passes  through a nonlinear High-Power-Amplifier (HPA). The non-linearity of HAP leads to in-band distortion which increases Bit Error Rate (BER), and out-of-band radiation, which causes adjacent channel interference.

There are several proposals to deal with the PAPR problem in OFDM systems \cite{overview1}, \cite{overview2}. The different techniques can be classified into different groups according to their characteristics. The most general classification is: clipping techniques, coding techniques, the distortionless schemes with side information and distortionless techniques without side information. 

The simplest implementation method is clipping technique, which consists in to deliberately clip the OFDM signal before amplification \cite{Ochiai_clip}. Clipping can reduce PAPR but this is a nonlinear process and may cause both in-band and out-of-band interference while destroying the orthogonality among the subcarriers.  Then, coding techniques are found, which are introduced in \cite{Jones-coding}. The key of those techniques is to select the codewords that minimize the PAPR.  In the next group we have the techniques that cause no distortion and create no out-of-band radiation, but they may be require the transmission of the side information to the receiver.  Techniques that require the transmission of side information are for example Partial Transmit Sequence (PTS),  Tone Reservation  (TR), etc. On the other hand SeLected Mapping (SLM), Constellation Extension or Orthogonal Pilot Sequences (OPS) do not require the transmission of side information.

In this paper we describe some relevant PAPR reduction techniques of the literature. Therefore, the paper is organized as follows. Section \ref{sec:signal-model} briefly shows the OFDM signal model. In Section \ref{sec:papr} the PAPR problem of the OFDM system is presented. The Clipping techniques are exposed in Section \ref{sec:clip}. In Section \ref{sec:coding} coding schemes are described. The analysis of the different distortionless techniques are addressed in Section \ref{sec:dstless}. Simulation results are provided in Section \ref{sec:resul}. Finally conclusions are drawn in Section \ref{sec:viii}.

\section{The OFDM  Signal Model}
\label{sec:signal-model}

\begin{figure}[h] 
\centering
\includegraphics[width=0.51\textwidth]{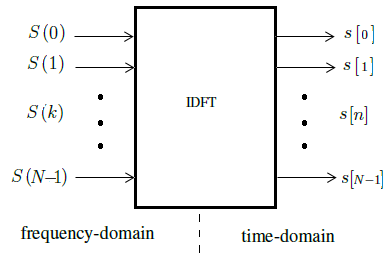}
\caption{The IDFT implementation of an OFDM symbol.}
\label{fig:ofdm-signal}
\end{figure}

The OFDM signal is the sum of $N$ independent signals modulated onto subchannels of equal bandwidth, which can be efficiently implemented by an Inverse Discrete Fourier Transform (IDFT) operation, as illustrated in Figure \ref{fig:ofdm-signal}. 

We denote $\mathbf{S}^{\ell}=\lbrace S^{\ell}(k)\rbrace$ as the frequency-domain complex data sequence, consisting of the frequency-domain complex symbols over $k$th subcarrier, $k=\{0,\cdots, N-1\}$ to be transmitted in the $\ell$th OFDM symbol. Then the time-domain signal $\mathbf{s}^{\ell}=\lbrace s^{\ell}\left[n\right] \rbrace$ is given by:

\begin{equation}\label{ofdm}
s^{\ell}\left[n\right]=\frac{1}{\sqrt{N}} \sum^{N-1}_{k=0} {S^{\ell}(k) e^{j \frac{2\pi}{N} kn } }\, , 0\leq n < N - 1
\end{equation} where $k$ and $n$ are the frequency and time indices respectively. For the convenience, we omit the use of superscript $\ell$ in the sequel.

\section{The PAPR Problem in OFDM Systems}
\label{sec:papr}

The frequency-domain data sequence $\mathbf{S}={\lbrace S(k) \rbrace}$ are independent, identically distributed (i.i.d.) random variables and due to the central limit theorem, a small percentage of output samples will take very large magnitudes. This results in the well-known PAPR problem of OFDM systems.

In general, the PAPR ($\chi$) of the time-domain sequence $ \mathbf{s}=\lbrace s\left[n\right] \rbrace$ is defined as the ratio between the maximum instantaneous power and its average power \cite{tellado}, that is: 

\begin{equation}\label{papr}
\chi=PAPR \lbrace \mathbf{s} \rbrace = \frac{\mbox{max} \left( {\left| {\mathbf{s}} \right|}^2 \right)} {E\left\{ {\left| \mathbf{s} \right|}^2\right\}}
\end{equation} where $E{\left\{ \cdot \right\}}$ denotes expected value. 

In the literature, the most common way to evaluate the PAPR is to determine the probability that this PAPR exceeds a certain threshold $\chi_0$. This is represented by the Complementary Cumulative Distribution Function (CCDF), which is a random variable, as:

\begin{equation}\label{ccdf}
CCDF(\chi) = Prob(\chi^{\ell} > \chi_0)=1-1(1-e^{\chi_0})^N
\end{equation} 

\section{Clipping techniques}
\label{sec:clip}

The simplest PAPR reduction method consists basically in clipping the high parts of the signal amplitude that are outside the allowed region \cite{Ochiai_clip}. If the OFDM symbol $\mathbf{s}$ is clipped at a level $A$, then the the clipped signal $\tilde{\mathbf{s}}$ is:

\begin{equation}
\label{eq:clip}
\tilde{\mathbf{s}} = \left\{ 
	\begin{array}{lcr}
		A, 			                & \vert \mathbf{s}  \vert \leq A\\
		Ae^{j \phi(\mathbf{s}) },   & \vert \mathbf{s}  \vert >  A
	\end{array} 
	\right.
\end{equation} where $\phi(\mathbf{s})$ is the phase of $\mathbf{s}$.

This technique is the simplest of implementation but it has the following drawbacks:

\begin{itemize}
\item Clipping causes in-band distortion, which degrades the performance of the BER

\item Clipping causes out-of-band radiation, resulting in adjacent interference. This can be reduced by filtering,  and thus Clipping and Filtering (CF) operation is used in \cite{Armstrong-clip-filtering}. When the signal passes through the low-pass filtering there is a peak power regrowth. In \cite{Ochiai_clip} it has been shown that the Nyquist-rate clipping suffers from a much higher peak power regrowth compared to the clipping with oversampling. Thus, the results suggest that for efficient reduction of the peak power, the OFDM should be sufficiently oversampled (\textit{i.e.} $L\geq 4$) before clipping.

\item There is the possibility to use iterative CF \cite{Deng-R-clip-filtering}, but it takes many iterations to reach a desired amplitude level $A$.

\end{itemize}

\section{Coding techniques}
\label{sec:coding}

Coding techniques consist in selecting the codewords that minimize or reduce the PAPR. Initially the idea was introduced in \cite{Jones-coding}. This scheme requires exhaustive computational load to search the best codewords and to store the large lookup tables for encoding and decoding, specially with large number of subcarriers. Moreover, these techniques do not address the error correction problem. In \cite{Jones-coding2}, an optimum code set for achieving minimum PAPR is introduced and, moreover, to deal with the error correction it uses an additive offset. It enjoys the twin benefits of power control and error correction, but requires extensive calculation to find good codes and offsets.

Also, there are approaches in which the use of Complementary Block Coding (CBC) was proposed to reduce the PAPR  without the restriction of the frame size \cite{Davis-Golay}, \cite{Jiang-cbc}. In \cite{Fischer-RS} Reed-Solomon (RS) codes over the Galois field are employed to create a number of candidates, from which the best are selected.  Considering the characteristics of those coding techniques, the main disadvantage of those coding methods is the good performance at the cost of coding rate, and a high computational load to search the adequate codewords.

\section{Distortionless techniques}
\label{sec:dstless}

As an alternative to combat the PAPR problem, there are several techniques without distortion such us Partial Transmit Sequences (PTS), SeLected Mapping (SLM), Tone Reservation (TR), Orthogonal Pilot Sequences (OPS) and Constellation Extension among other. Those methods may require the transmission of side information to the receiver. In the next lines we dedicate to explain  the several distortionless techniques put into two groups according to if they require or not the transmission of side information.

\subsection{With Side Information}

Many techniques require the transmission of side information to the receiver in order to the receiver can  determine the type of processing that has been employed at the transmitter side. We introduced the more significant schemes.

\subsubsection{Partial Transmit Sequences (PTS)} The PTS technique was originally introduced in \cite{muller-pts}, which consists in that the frequency-domain input data block $\mathbf{S}=\lbrace S(k)\rbrace$ is subdivided into $V$ disjoint carrier subblocks, which are represented by the vectors $\lbrace \mathbf{S}^{(v)}, \quad v=0,\cdots,V-1 \rbrace$. 

In general, for PTS scheme, the known subblock partitioning methods can be classified into three categories \cite{muller-pts}: adjacent partition, interleaved partition and pseudorandom partition. Then, the subblocks $\mathbf{S}^{(v)}$ are transformed into $V$ time-domain partial transmit sequences:

\begin{equation}
\label{eq:pts-td}
\mathbf{s}^{v}=\left[ s_{0}^{v}, s_1^{v}, \cdots, s_{N-1}^{v} \right] = \mbox{IDFT} \lbrace{\mathbf{S}}^{(v)}\rbrace.
\end{equation}

These partial sequences are independently rotated by phase factors $\mathbf{b}=\lbrace b_v= e^{j\theta_v}\rbrace$. The objective is to optimally combine the subblocks to obtain the time-domain OFDM signals with the lowest PAPR.

\begin{equation}
\label{eq:pts}
\tilde{\mathbf{s}}= \sum_{v=0}^{V-1}{ b_v } {\mathbf{s}^v}
\end{equation}

Therefore, there are two important issues that should be solved in PTS technique: high computational complexity to reach optimal phase factors and the overhead of the phase factors as side information needed to be transmitted to the receiver for correct decoding. 

Recently,  some approaches have been proposed in order to reduce the complexity, such us \cite{cimini-pts}, where an iterative PTS (IPTS) is presented. The performance of this scheme is not good although the IPTS is very simple.  The dual-layer phase sequencing (DLPS) with different implementations also is proposed in \cite{ho-pts}. There are other techniques that have been introduced in \cite{yang-pts}, \cite{lim-pts} and \cite{ghassemi-pts} where the authors propose to reduce the computational complexity of PTS.  For example, in \cite{yang-pts} is introduced the idea of the relationship between the weighting factors and the transmitted bit vectors.

\subsubsection{Tone Reservation (TR) and Tone Injection (TI)} TR and TI are two efficient techniques to reduce the PAPR of OFDM signals and they are proposed in \cite{tellado}. In these schemes both transmitter and receiver reserve a subset of tones that are not used for data transmission to generate PAPR reduction signals.

The objective of TR is to find the time-domain signal $\mathbf{c}$ to be added to the original time-domain signal $\mathbf{s}$ to reduce the PAPR. Denoting $\mathcal{C}=\lbrace C_0, \cdots, C_{N-1} \rbrace $ as a frequency-domain vector for tone reservation, that added to the data input symbols $\mathbf{S}$, then the new time-domain signal after tone reservation processing is $\tilde{\mathbf{s}}$:

\begin{equation}
\label{tr-td}
\tilde{\mathbf{s}}= \mbox{IDFT} \lbrace  \mathbf{S} + \mathcal{C}\rbrace
\end{equation} 

Therefore, the main aim of the TR is to find out the proper $\mathbf{c}$ to make the vector with lower PAPR. To find the value of $\mathcal{C}$, we must solve a convex optimization problem that can easily be cast as a linear programming problem.

Similarly, TI also uses an additive correction $\mathcal{C}$ \cite{ti}, although the idea is to increase the constellation size so that each of the points in the original basic constellation can be mapped into several equivalent points in the expanded constellation. Each symbol in a data block can be mapped into one of several equivalent constellation points. This method is called tone injection because substituting a point in the basic constellation for a new point in the larger constellation is equivalent to injecting an appropriate tone.

The TI technique is more problematic than the TR scheme since the injected signal occupies the frequency band as the information bearing signals. Moreover, the alternative constellation points in TI technique have an increased energy and the implementation complexity increases due to the computation of the
optimal translation vector \cite{overview2}.
 
\subsection{Without Side Information}

In the next lines of this subsection we expose several distortionless techniques, which do not require the transmission of side information to the receiver.

\subsubsection{SeLected Mapping (SLM)} In this technique, the transmitter generates a sufficiently large number of alternative OFDM signal sequences, all representing the same information as the original symbol. Then each of these alternative input data sequences is made the IDFT operation and the one with the lowest PAPR is selected for transmission \cite{huber-slm}.

Each input data symbol $\mathbf{S}$ is multiplied by $U$ different phase sequences, each of length $N$, $\mathbf{B}_{(u)}=\lbrace b_{(u, 0)}, \cdots, b_{(u, N-1)} \rbrace,\quad u=\lbrace 0, \cdots, U-1 \rbrace$, resulting in $U$ modified data symbol. After applying SLM to $\mathbf{S}$, the time-domain signal $\mathbf{s}^{u}$ becomes:

\begin{equation}
\label{slm/td}
	\mathbf{s}^{u} = \frac{1}{\sqrt{N}} \sum_{k=0}^{N-1} S(k)b_{(u,k)} e^{j\frac{2\pi}{N}kn}, 
\end{equation} where $0\leq n<N-1,\quad 0\leq u<U-1$.

The original SLM scheme has the next characteristics:

\begin{itemize}
\item Information about the selected phase sequence should be transmitted to the receiver as side information. At the receiver, the reverse operation is performed to recover the original data symbol. However, in \cite{muller-slm} and \cite{legoff-slm} an SLM algorithm without explicit side information is proposed.

\item For implementation, the SLM technique requires a bank of $U$ IDFT operations to generate a set of candidate transmission signals, and this requirement usually results in high
computational complexity. There are some approaches attempting to decrease the complexity like \cite{wang-slm}, \cite{heo-slm} and \cite{ghassemi-slm}.

\item This approach is applicable with all types of modulation and any number of subcarriers. \item The amount of PAPR reduction for SLM depends on the number of phase sequences $U$ and the design of the phase sequences.
\end{itemize}

\subsubsection{Constellation Extension}  
\begin{figure}[h] 
\centering
\includegraphics[width=0.30\textwidth]{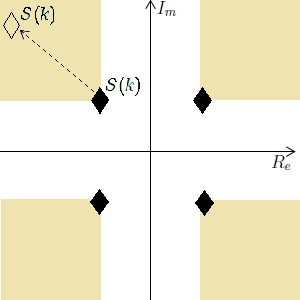}
\caption{Extension of the outer constellation points with QPSK encoding. The
shaded region represents the extension region.}
\label{fig:constellation}
\end{figure}

In Constellation Extension techniques the key is to play intelligently with the outer constellation points, that are moved within the proper quater-plane as shown in Figure \ref{fig:constellation}, such that the PAPR is minimized.

The main advantages of these techniques are enumerated next:
\begin{itemize}
\item  The minimum distance of the constellation points is not affected and it consequently guarantees no BER degradation.
\item These methods do no require the transmission of side information to the receiver.
\item There is no data rate loss.  
\end{itemize}

Nevertheless, constellation expansion schemes introduce an increase in the energy per symbol.

In \cite{ace} Active Constellation Extension (ACE) is presented, and in this method all symbols are expanded but its computational burden is high.  More recent, a metric-based symbol predistortion scheme has been introduced in \cite{sap1} and \cite{sap2}. In this method, a metric, defined mathematically in (\ref{eq:metrica}), is used to measure how much each frequency-domain symbol contributes to large peaks, and the frequency-domain symbols with the highest metric values are selected and predistorted with a constant scaling factor $\alpha$. 

\begin{equation}
\label{eq:metrica}
\mu_k = \sum_{{n \in T_K}} \omega(n)f(n,k)
\end{equation} where, $f(n,k)=-\cos(\varphi_{nk})$ is a function which gives an appropriate measure of the phase angle between the output sample $s[n]$ and the contribution of symbol $S(k)$, and $\varphi_{nk}$ is the angle between the output sample and the contribution of the symbol, $\omega(n)={\left | \mathbf{s} \right |}^p$ is a weighting function that gives more weight to the output samples with large magnitudes, $p$ is a selected parameter, and $T_K$ is a set of size $K$ whose elements are the indices of the output samples that are larger than a predetermined threshold value \cite{sap1}.  Then the $L$ symbols with greatest positive metrics are determined to be predistorted with the corresponding scaling factor $\alpha>1$. 

This metric-based algorithm saves energy (since only a subset of symbols are amplitude predistorted) and it avoids a high computational load with the use of a metric. However, the main drawback is that the size of the subset and the scaling factor are chosen from a group of values suggested by the authors after empirical search.

\subsubsection{Orthogonal Pilot Sequences} In coherent wireless OFDM systems, pilot symbols are usually inserted in the 2D time-frequency grid to estimate the channel. OPS technique is proposed in \cite{fernandez-ops}, which inserts the orthogonal pilot sequences in the input symbols that provides the lowest PAPR. 

In a OFDM symbol with $N$ subcarriers, a subset $\Upsilon$ of subcarriers will carry pilot symbols and thus, the input data symbols $\mathbf{S}={\lbrace S(k)\rbrace}^{N-1}_{k=0}$ are:

\begin{equation}
\label{eq:ops-simbolos}
S\left(k\right) = \left\{ 
	\begin{array}{lcr}
		X(k), & k\notin\Upsilon\\
		P(k), & k\in\Upsilon 		
	\end{array} 
\right.
\end{equation} where $P(k)$ and $X(k)$ are pilot and data symbols respectively. 

The transmitted time-domain symbol $s[n]=x[n]+p[n]$ can be separated into two parts, as:

\begin{equation}
\label{eq:ops-td}
	s\left[n\right] = \left\{ 
	\begin{array}{lcr}
		x\left[n\right] &=\frac{1}{\sqrt{N}}\sum_{k\notin\Upsilon}{X(k)e^{j \frac{2\pi}{N}kn}} \\
		p\left[n\right] &=\frac{1}{\sqrt{N}}\sum_{k\in\Upsilon}{P(k)e^{j \frac{2\pi}{N}kn}}
	\end{array}
\right.
\end{equation} where $x\left[n\right]$ and $p\left[n\right] $ refer to the time-domain data and pilot signals, respectively. 

OPS technique \cite{fernandez-ops} proposes the use of a predetermined set of $M$ orthogonal pilot sequences of length $N_p$ ($M \leq N_p$) in order to reduces complexity and avoids any side information since blind detection is possible at the receiver due to the orthogonality condition. 

The $N_p$ pilot symbols of each OFDM symbol can be collected in a $N-$length sequence denoted as $P$ where, the $k$th element of this sequence is given by:

\begin{equation}\label{pilotos}
	{ [ P ]}_k= \left\{ 
		\begin{array}{lr}
			P(k), & k \in\Upsilon \\
			0	  & k \notin\Upsilon
		\end{array}
	\right.
\end{equation}

As stated before, a set of $M$ pilot sequences are available so the alphabet of $P$ is $\{P_1, P_2, \cdots , P_M\}$. Each pilot sequence of this finite set $P_m$, $m \in \{ 1, 2, \cdots,M\}$ contains the frequency-domain pilot symbols at pilot positions while zeros are inserted in the remaining ones. These pilots sequences are orthogonal so then the ortoghonality conditions is fulfilled

\begin{equation}\label{orthog}
\begin{array}{lcc}
\langle P_m, P_n \rangle = 0 & m \neq n & m,n=\{1, \cdots, M\}
\end{array}
\end{equation}where $\langle \cdot,\cdot \rangle$ denotes the inner product.

In particular, if the well-known Walsh-Hadamard sequences are employed where $P^{\ell}(k) \in \{ 1,-1 \}$, then $\langle P_m, P_n \rangle = N_p  \delta [m-n] $, $ m,n=\{1, \cdots, M\}$, where $\delta[\cdot]$ denotes the Kronecker delta function. 

\section{Simulation Results}
\label{sec:resul}

Simulations are presented by averaging over $10^4$ randomly OFDM symbols with Quaternary Phase-Shift Keying (QPSK) modulation. The performance of the several PAPR schemes is presented in terms of CCDF.

\begin{figure}[h] 
\centering
\includegraphics[width=0.49\textwidth]{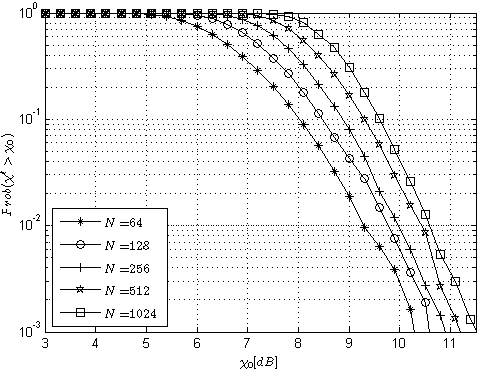}
\caption{CCDF of the OFDM symbols without any PAPR technique for $N=\lbrace 64, 128, 256, 512, 1024 \rbrace$ subcarriers.}
\label{fig:ccdf-Ns}
\end{figure}

Figure \ref{fig:ccdf-Ns} illustrates the PAPR of an OFDM system before applying any PAPR reduction technique. We present the CCDF for a set of different values of $N=\lbrace 64, 128, 256, 512, 1024 \rbrace$ subcarriers. The horizontal and vertical axis represent the threshold ($\chi_0$) for the PAPR and the probability that the PAPR of a certain OFDM symbol exceeds a threshold, respectively. It is shown that the unmodified OFDM signal has a PAPR that exceeds 10.5 dB at the probability $10^{-3}$ for $N = 256$.
 
\begin{figure}[h] 
\centering
\includegraphics[width=0.51\textwidth]{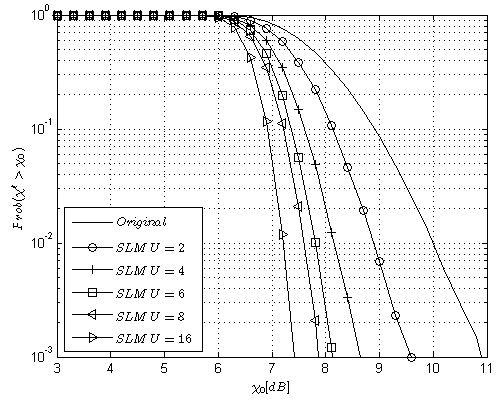}
\caption{CCDF of PAPR  for $N=256$ subcarriers.  The solid-line curve corresponds to the conventional OFDM signal without any PAPR reduction scheme. The solid marked line curves represent the SLM technique with different values of $U$}
\label{fig:ccdf-slm}
\end{figure}

Figure \ref{fig:ccdf-slm} shows the performance comparison in terms of PAPR reduction with SLM technique when we employ  different values of the number of phase sequences $U$. The solid line curve represents the CCDF of the OFDM symbols without any PAPR technique, and the solid marked line curves show the performance of the OFDM symbols after applying SLM method for different values of $U=\lbrace 2, 4, 6, 8, 16 \rbrace$. It can observed how a reduction close to 1.5 dB with $U=2$ at a probability of $10^{-3}$.

\begin{figure}[h] 
\centering
\includegraphics[width=0.51\textwidth]{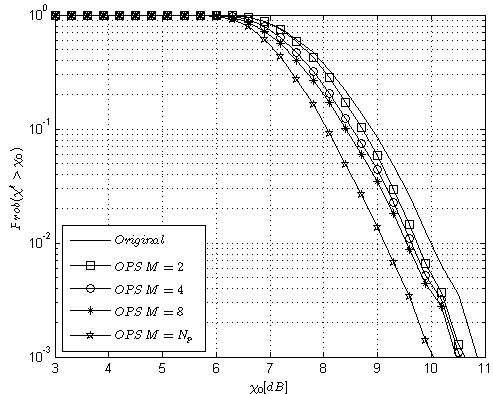}
\caption{CCDF of PAPR  for $N=\lbrace 256 \rbrace$ subcarriers.  The solid-line curve corresponds to the conventional OFDM signal without any PAPR reduction scheme. The solid marked line curves represent the OPS technique with different values of $M$.}
\label{fig:ccdf-ops}
\end{figure}

Simulations in Figure \ref{fig:ccdf-ops} depict the CCDF of the PAPR reduction after applying OPS technique with different values of $M$. We employ $N=256$ subcarriers. The CCDF of the OFDM symbols without any PAPR technique is represented by the solid line curve and the solid marked line curves show the performance of OPS scheme for different values of $M=\lbrace 4, 8, 16 \rbrace$. It can observed how the reduction is in the order of 1.5 dB with $U=2$ at a probability of $10^{-3}$.

\begin{figure}[h] 
\centering
\includegraphics[width=0.50\textwidth]{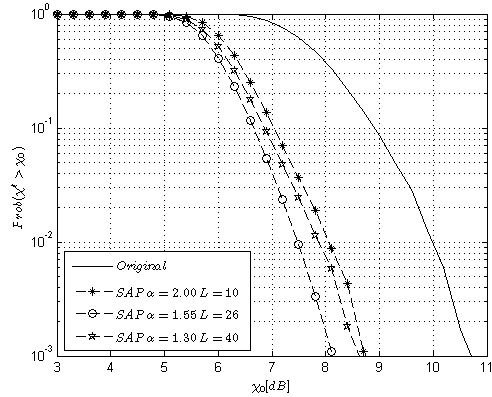}
\caption{PAPR performance for $N=256$ subcarriers.  The solid-line curve corresponds to the conventional OFDM signal without any PAPR reduction scheme. The solid marked line curves represent the SAP technique with different set values of $\alpha \, \mbox{and} \, L $.}
\label{fig:ccdf-sap}
\end{figure}

Results of simulations of Simple Amplitude Presitortion (SAP) technique \cite{sap1} are presented in Figure \ref{fig:ccdf-sap}. We employ $N=256$ subcarriers. The CCDF of the OFDM symbols without any PAPR technique is represented by the solid line curve and the solid marked line curves illustrate the performance of SAP scheme for different values of $\alpha$ and $L$. It can be noticed how the reduction is close to 2.5 dB with $\alpha=1.55$ at a probability of $10^{-3}$.

\section{Conclusion}
\label{sec:viii}

OFDM is an attractive technique for digital communications systems due to its high bit rate, strong immunity to multipath and high spectral efficiency. However, one of the most serious problems is the high Peak-to-Average Power Ratio (PAPR) of the transmitted OFDM signal, since this large peaks introduce a serious degradation in performance when the signal passes through a nonlinear High-Power-Amplifier (HPA). In this paper, we address the PAPR problem of OFDM systems and the more relevant techniques to achieve PAPR reduction. The more remarkable characteristics of those techniques are discussed as well as it is provided their mathematical description although there is an extensive state of the art, nowadays, the PAPR problem is still an active area of research with many open issues.

%
\section*{Acknowledgment}
This work has been partly funded by the Spanish national projects GRE3N-SYST (TEC2011-29006-C03-03) and COMONSENS (CSD2008-00010), and SENESCYT (Ecuador).

\ifCLASSOPTIONcaptionsoff
  \newpage
\fi



\begin{biography}[{\includegraphics[width=1in,height=1.24in]{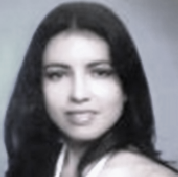}}]{Martha C.~Paredes Paredes} 
received the Ingeniero en Electr\'{o}nica y Redes de Informaci\'{o}n degree from Escuela Polit\'{e}cnica Nacional, Quito, Ecuador in 2008 and the M.~Sc.~of Multimedia and Communications from Carlos III University of Madrid, Spain in 2010, with scholarship from Fundaci\'{o}n Carolina, Spain. She is currently pursuing the Ph.D.~degree in the Department of Signal Theory and Communications at Carlos III University of Madrid, where she is doing research on signal processing for multicarrier modulation and PAPR reduction in OFDM systems.
\end{biography}

\begin{biography}[{\includegraphics[width=1in,height=1.25in]{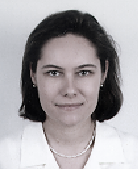}}]
{M.~Julia Fern\'{a}ndez-Getino Garc\'{i}a} 
received the M.~Eng. and Ph.D. degrees in telecommunication engineering, both from the Polytechnic University of Madrid, Spain, in 1996 and 2001, respectively. Currently, she is with the Department of Signal Theory and Communications of Carlos III University of Madrid, Spain, as an Associate Professor. From 1996 to 2001, she held a research position at the Department of Signals, Systems and Radiocommunications of Polytechnic University of Madrid. She was on leave during 1998 at Bell Laboratories, Murray Hill, NJ, USA, visited Lund University, Sweden, during two periods in 1999 and 2000, visited Politecnico di Torino, Italy, in 2003 and 2004, and visited Aveiro University, Portugal, in 2009 and 2010. Her research interests include multicarrier communications, coding and signal processing for wireless systems. In 1998 and 2003, she respectively received the best `Master Thesis' and `Ph.D. Thesis' awards from the Professional Association of Telecommunication Engineers of Spain, and in 1999 and 2000, she was respectively, awarded the `Student Paper Award' and `Certificate of Appreciation' at the IEEE International Conferences PIMRC'99 and VTC'00. In 2004, she was distinguished with the `Ph.D. Extraordinary Award' from the Polytechnic University of Madrid. In 2012, she has received the `Excellence Award' to her research from Carlos III University of Madrid.
\end{biography}


\end{document}